\documentclass[12pt, centerh1]{article}
\usepackage{lipsum}
\usepackage{mathtools}

\newcommand{\btheta}{{\boldsymbol{\theta}}}
\newcommand{\bvtheta}{{\boldsymbol{\vartheta}}}

\newcommand\numberthis{\addtocounter{equation}{1}\tag{\theequation}}

\newcommand{\vecmu}{\mbox{\boldmath$\mu$}}

\newcommand{\vecx}{\mathbf{x}}

\newcommand{\vecX}{\mathbf{X}}

\newcommand{\vecV}{\mathbf{V}}

\newcommand{\matsig}{\mathbf\Sigma}
\newcommand{\matPsi}{\mathbf\Psi}

\newcommand{\tr}{\,\mbox{tr}}

\newcommand{\vecA}{\mathbf{A}}

\newcommand{\matm}{\mathbf{M}}

\newcommand{\vecalp}{\boldsymbol{\alpha}}
\newcommand{\vecc}{\text{vec}}
\newcommand{\fX}{\mathscr{X}}
\newcommand{\fV}{\mathscr{V}}
\newcommand{\tgamma}{\tilde{\gamma}}
\newcommand{\vecT}{\mathbf{T}}

\oddsidemargin=0mm\usepackage{natbib}
\usepackage{amsmath,natbib}
\usepackage{amsfonts,mathrsfs,moreverb,lipsum}
\usepackage{graphicx,colonequals}
\usepackage{fixltx2e}
\usepackage{color}
\usepackage{caption}
\usepackage{subcaption}
\usepackage{pbox}
\title{Three Skewed Matrix Variate Distributions}
\author{Michael P.B.\ Gallaugher and Paul D.\ McNicholas}
\date{\small Dept.\ of Mathematics \& Statistics, McMaster University, Hamilton, Ontario, Canada.}

\begin{document}
\maketitle
\begin{abstract}
Three-way data can be conveniently modelled by using matrix variate distributions. Although there has been a lot of work for the matrix variate normal distribution, there is little work in the area of matrix skew distributions. Three matrix variate distributions that incorporate skewness, as well as other flexible properties such as concentration, are discussed. Equivalences to multivariate analogues are presented, and moment generating functions are derived. Maximum likelihood parameter estimation is discussed, and simulated data is used for illustration. 
\end{abstract}

\section{Introduction}
Matrix variate distributions are useful in modelling three way data, e.g., multivariate longitudinal data. Although the matrix normal distribution is widely used, there is relative paucity in the area of matrix skewed distributions. Herein, we discuss matrix variate extensions of three already well established multivariate distributions using matrix normal variance-mean mixtures. Specifically, we consider a matrix variate generalized hyperbolic distribution, a matrix variate variance-gamma distribution, and a matrix variate normal inverse Gaussian (NIG) distribution. Along with the matrix variate skew-$t$ distribution, mixtures of these respective distributions have been used for clustering \citep{gallaugher18}; however, unlike the matrix variate skew-$t$ distribution \citep{gallaugher17a}, their properties have yet to be discussed and this letter aims to fill that gap.

\section{Background}
\subsection{The Matrix Variate Normal and Related Distributions}
One of the most mathematically tractable examples of a matrix variate distribution is the matrix variate normal distribution. An $n\times p$ random matrix $\fX$ follows a matrix variate normal distribution if its probability density function can be written as
$$
f(\vecX|\matm,\matsig,\matPsi)=\frac{1}{(2\pi)^{\frac{np}{2}}|\matsig|^{\frac{p}{2}}|\matPsi|^{\frac{n}{2}}}\exp\left\{-\frac{1}{2}\tr\left(\matsig^{-1}(\vecX-\matm)\matPsi^{-1}(\vecX-\matm)'\right)\right\},
$$ 
where $\matm$ is an $n\times p$ location matrix, $\matsig$ is an $n\times n$ scale matrix for the rows of $\fX$ and $\matPsi$ is a $p\times p$ scale matrix for the columns of $\fX$. We denote this distribution by $\mathcal{N}_{n\times p}(\matm,\matsig,\matPsi)$ and, for notational clarity, we will denote the random matrix by $\fX$ and its realization by $\vecX$.
One useful property of the matrix variate normal distribution, as given in \cite{harrar08}, is 
\begin{equation}
\fX\sim\mathcal{N}_{n\times p}(\matm,\matsig,\matPsi)\iff \vecc(\fX)\sim\mathcal{N}_{np}(\vecc(\matm),\matPsi\otimes\matsig),
\label{eq:normprop}
\end{equation}
where $\mathcal{N}_{np}(\cdot)$ denotes the multivariate normal distribution with dimension $np$, $\vecc(\cdot)$ denotes the vectorization operator, and $\otimes$ denotes the Kronecker product. 

Although the matrix variate normal is probably the most well known matrix variate distribution, there are other examples. For example, the Wishart distribution \citep{Wishart} was shown to be the distribution of the sample covariance matrix for a random sample from a multivariate normal distribution. There are also a few examples of a matrix variate skew normal distribution such as \cite{chen2005}, \cite{dominguez2007} and \cite{harrar08}. Most recently, \cite{gallaugher17a}, considered a matrix variate skew-$t$ distribution using a matrix normal variance-mean mixture.

There are also a few examples of mixtures of matrix variate distributions. \cite{Anderlucci15} considered a mixture of matrix variate normal distributions for clustering multivariate longitudinal data and \cite{dougru16} considered a mixture of matrix variate $t$ distributions. 

\subsection{The Inverse and Generalized Inverse Gaussian Distributions}
The derivation of the matrix distributions and parameter estimation discussed in Section~3, will rely heavily on the generalized inverse Gaussian distribution, and to a lesser extent the inverse Gaussian distribution.
A random variable $Y$ follows an inverse Gaussian distribution if its probability density function is of the form
$$
f(y|\delta,\gamma)=\frac{\delta}{\sqrt{2\pi}}\exp\{\delta\gamma\}y^{-\frac{3}{2}}\exp\left\{-\frac{1}{2}\left(\frac{\delta^2}{y}+\gamma^2y\right)\right\}
$$
for $y>0$, where $\delta,\gamma>0$.
For notational purposes, we will denote this distribution by $\text{IG}(\delta,\gamma)$.

The generalized inverse Gaussian distribution has two different parameterizations, both of which will be useful. A random variable $Y$ has a generalized inverse Gaussian distribution parameterized by $a, b >0$ and $\lambda\in \mathbb{R}$, denoted by $\text{GIG}(a,b,\lambda)$, if its probability density function can be written as
$$
f(y|a, b, \lambda)=\frac{\left({a}/{b}\right)^{\frac{\lambda}{2}}y^{\lambda-1}}{2K_{\lambda}(\sqrt{ab})}\exp\left\{-\frac{ay+{b}/{y}}{2}\right\}
$$
for $y>0$, where
$$
K_{\lambda}(u)=\frac{1}{2}\int_{0}^{\infty}z^{\lambda-1}\exp\left\{-\frac{u}{2}\left(z+\frac{1}{z}\right)\right\}dz
$$
is the modified Bessel function of the third kind with index $\lambda$. 
Some expectations of functions of a GIG random variable with this parameterization have a mathematically tractable form, e.g.,
\begin{equation*}\begin{split}
&\mathbb{E}(Y)=\sqrt{\frac{b}{a}}\frac{K_{\lambda+1}(\sqrt{ab})}{K_{\lambda}(\sqrt{ab})},
\qquad
\mathbb{E}\left({1}/{Y}\right)=\sqrt{\frac{a}{b}}\frac{K_{\lambda+1}(\sqrt{ab})}{K_{\lambda}(\sqrt{ab})}-\frac{2\lambda}{b},\\
&\mathbb{E}(\log Y)=\log\left(\sqrt{\frac{b}{a}}\right)+\frac{1}{K_{\lambda}(\sqrt{ab})}\frac{\partial}{\partial \lambda}K_{\lambda}(\sqrt{ab}).
\end{split}\end{equation*}

Although this parameterization of the GIG distribution will be useful for parameter estimation, for the purposes of deriving the density of the matrix variate generalized hyperbolic distribution, it is more useful to take the parameterization
\begin{equation}
g(y|\omega,\eta,\lambda)= \frac{\left({w}/{\eta}\right)^{\lambda-1}}{2\eta K_{\lambda}(\omega)}\exp\left\{-\frac{\omega}{2}\left(\frac{w}{\eta}+\frac{\eta}{w}\right)\right\},
\label{eq:I}
\end{equation}
where $\omega=\sqrt{ab}$ and $\eta=\sqrt{a/b}$. For notational clarity, we will denote the parameterization given in \eqref{eq:I} by $\text{I}(\omega,\eta,\lambda)$.

\subsection{Variance-Mean Mixtures}
A $p$-variate random vector $\vecX$ defined in terms of a variance-mean mixture, has a probability density function of the form
$$
f(\vecx)=\int_{0}^{\infty}\phi_{p}(\vecx|\vecmu+w\vecalp,w\matsig)h(w|\btheta)dw,
$$
where the random variable $W>0$ has density function $h(w|\btheta)$, and $\phi_{p}(\cdot)$ represents the density function of the $p$-variate Gaussian distribution. This representation is equivalent to writing
\begin{equation}
\vecX=\vecmu+W\vecalp+\sqrt{W}\vecV,
\label{eq:mvmixture}
\end{equation}
where $\vecmu$ is a location parameter, $\vecalp$ is the skewness, $\vecV\sim \mathcal{N}_{p}({\bf 0},\matsig)$ with $\matsig$ as the scale matrix, and $W$ has density function $h(w|\btheta)$. Note that $W$ and $\vecV$ are independent.
Many multivariate distributions can be obtained through a variance mean mixture by changing the distribution of $W$. 
For example, the $p$-dimensional generalized hyperbolic distribution, $\text{GH}_p(\vecmu,\vecalp,\matsig,\psi,\chi,\lambda)$,  as given in \cite{mcneil05}, was shown to arise as a special case of \eqref{eq:mvmixture} by taking $W\sim\text{GIG}(\psi,\chi,\lambda)$. However, there was a restriction that $|\matsig|=1$. Simply relaxing this constraint results in an identifiability problem. In \cite{browne15}, this was discussed, and the authors proposed the reparameterization $\omega=\sqrt{\psi\chi}, \eta=\sqrt{{\chi}/{\psi}}$. The representation of $\vecX$ is then as in \eqref{eq:mvmixture}, with  $W\sim \text{I}(\omega,1,\lambda)$. 

The $p$-dimensional variance-gamma distribution, $\text{VG}_p(\vecmu,\vecalp,\matsig,\lambda,\psi)$, results as a limiting case of the generalized hyperbolic by taking $\lambda>0$, and $\chi\rightarrow 0$. The precise details can be found in \cite{mcnicholas17}; in essence, the variance-gamma distribution also arises as a special case of \eqref{eq:mvmixture}, with $W\sim \text{gamma}(\lambda,\psi/2)$, where $\text{gamma}(a,b)$ denotes the gamma distribution with density 
$$
f(w|a,b)=\frac{b^a}{\Gamma(a)}w^{a-1}\exp\{-bw\}
$$
for $w>0$, where $a,b>0$.
However, we again have an identifiability issue using this representation if we remove the constraint $|\matsig|=1$. In \cite{mcnicholas17}, the authors propose setting $\mathbb{E}(W)=1$, resulting in the reparameterization $\gamma\coloneqq\lambda=\psi/2$.

Finally, we have the $p$-dimensional Gaussian distribution, $\text{NIG}_p(\vecmu,\vecalp,\matsig,\delta,\gamma)$. In \cite{karlis09}, the authors derived the $p$-dimensional NIG distribution using a variance-mean mixture with $W\sim \text{IG}(\delta,\gamma)$. However, there was once again a restriction on the determinant of $\matsig$. To remove this restriction and maintain identifiability, \cite{karlis09} set $\delta=1$ and $\tgamma=\gamma$. 

\section{Three Matrix Variate Skew Distributions}
\subsection{Matrix Normal Variance-Mean Mixture}
We now present densities for matrix variate versions of the generalized hyperbolic, variance-gamma and NIG distributions.
For all three of these distributions, we consider a matrix normal variance-mean mixture, where we can take the representation
\begin{equation}
\fX=\matm+W\vecA+\sqrt{W}\fV,
\label{eq:mmvmix}
\end{equation}
where $\fV\sim \mathcal{N}_{n\times p}({\bf 0}_{n\times p},\matsig,\matPsi)$ with ${\bf 0}_{n\times p}$ representing the $n\times p$ zero matrix, $\matm$ is an $n\times p$ location matrix, $\vecA$ is an $n\times p$ skewness matrix, and $W>0$ is a random variable with density $h(\btheta)$. We now derive three matrix variate distributions using this representation with different distributions for $W$.

\subsection{A Matrix Variate Generalized Hyperbolic Distribution}
We now derive the density of a matrix variate generalized hyperbolic distribution. In this case, to avoid the indentifiability issue discussed in \cite{browne15}, we take $W\sim \text{I}\left(\omega,1,\lambda\right)$, where $\omega$ is a concentration parameter and $\lambda$ is the index parameter.
It then follows that
$$
\fX|w\sim \mathcal{N}_{n\times p}\left(\matm+w\vecA,w\matsig,\matPsi\right)
$$
and thus the joint density of $\vecX$ and $W$ is
\begin{align*}
f&(\vecX,w|\bvtheta)=f(\vecX|w)f(w)
= \frac{w^{\lambda-\frac{np}{2}-1}}{(2\pi)^{\frac{np}{2}}| \matsig |^{\frac{p}{2}} |\matPsi |^{\frac{n}{2}}2K_{\lambda}(\omega)} \\
& \quad\qquad \times \exp\left\{-\frac{1}{2w}\left(\tr\left(\matsig^{-1}(\vecX-\matm-w\vecA)\matPsi^{-1}(\vecX-\matm-w\vecA)'\right)+\omega\right)-{\omega w}/{2}\right\}, \numberthis \label{eqn:joint}
\end{align*}
where $\bvtheta=(\matm,\vecA,\matsig,\matPsi,\omega,\lambda)$.

We note that the exponential term in \eqref{eqn:joint} can be written as
$$
\exp\left\{\tr(\matsig^{-1}(\vecX-\matm)\matPsi^{-1}\vecA')\right\} \times \exp\left\{-\frac{1}{2}\left[\frac{\delta(\vecX;\matm,\matsig,\matPsi)+\omega}{w}+w\left(\rho(\vecA,\matsig,\matPsi)+\omega\right)\right]\right\},
$$
where
$\delta(\vecX;\matm,\matsig,\matPsi)=\tr(\matsig^{-1}(\vecX-\matm)\matPsi^{-1}(\vecX-\matm)')$ and $\rho(\vecA,\matsig,\matPsi)=\tr(\matsig^{-1}\vecA\matPsi^{-1}\vecA')$.
Therefore, the marginal density of $\vecX$ is
\begin{align*}
f(\vecX)&=\int_{0}^{\infty}f(\vecX,w) dw
=\frac{1}{(2\pi)^{\frac{np}{2}}| \matsig |^{\frac{p}{2}} |\matPsi |^{\frac{n}{2}}K_{\lambda}(\omega)}\exp\left\{\tr(\matsig^{-1}(\vecX-\matm)\matPsi^{-1}\vecA')\right\} \\
& \hspace{0.25in} \times \frac{1}{2} \int_{0}^\infty w^{\lambda-\frac{np}{2}-1}\exp\left\{-\frac{1}{2}\left[\frac{\delta(\vecX;\matm,\matsig,\matPsi)+\omega}{w}+w\left(\rho(\vecA,\matsig,\matPsi)+\omega\right)\right]\right\}dw. \numberthis \label{eq:Xmarg1}
\end{align*}
Making the change of variables given by
$$
y=\frac{\sqrt{\rho(\vecA,\matsig,\matPsi)+\omega}}{\sqrt{\delta(\vecX;\matm,\matsig,\matPsi)+\omega}}w,
$$
\eqref{eq:Xmarg1} becomes
\begin{align*}
f_{\text{MVGH}}(\vecX|\bvtheta)=&\frac{\exp\left\{\tr(\matsig^{-1}(\vecX-\matm)\matPsi^{-1}\vecA') \right\}}{(2\pi)^{\frac{np}{2}}| \matsig |^{\frac{p}{2}} |\matPsi |^{\frac{n}{2}}K_{\lambda}(\omega)}  \left(\frac{\delta(\vecX;\matm,\matsig,\matPsi)+\omega}{\rho(\vecA,\matsig,\matPsi)+\omega}\right)^{\frac{\left(\lambda-\frac{np}{2}\right)}{2}} \\ & \times
 K_{\left(\lambda-{np}/{2}\right)}\left(\sqrt{\left[\rho(\vecA,\matsig,\matPsi)+\omega\right]\left[\delta(\vecX;\matm,\matsig,\matPsi)+\omega\right]}\right),
\end{align*}
where $\omega>0$ is a concentration parameter, and $\lambda\in \mathbb{R}$ is an index parameter.

We note that the density of $\fX$, as derived here, resembles that of \cite{browne15}, and we denote this distribution by $\text{MVGH}_{n\times p}(\matm,\vecA,\matsig,\matPsi,\lambda,\omega)$.
For the purposes of parameter estimation, note that the conditional density of $W$ is
\begin{align*}
f(w|&\vecX)=\frac{f(\vecX|w)f(w)}{f(\vecX)}\\
&=\left(\frac{\rho(\vecA,\matsig,\matPsi)+\omega}{\delta(\vecX;\matm,\matsig,\matPsi)+\omega}\right)^{\frac{\left(\lambda-{np}/{2}\right)}{2}}\frac{w^{\lambda-{np}/{2}-1}}{2K_{\left(\lambda-{np}/{2}\right)}\sqrt{\left[\rho(\vecA,\matsig,\matPsi)+\omega\right]\left[\delta(\vecX;\matm,\matsig,\matPsi)+\omega\right]}}\\&\qquad\qquad\qquad\times\exp\left\{-\frac{(\rho(\vecA,\matsig,\matPsi)+\omega)w+{[\delta(\vecX;\matm,\matsig,\matPsi)+\omega]}/{w}}{2}\right\}.
\end{align*}
Therefore, $W|\vecX\sim \text{GIG}\left(\rho(\vecA,\matsig,\matPsi)+\omega,\delta(\vecX;\matm,\matsig,\matPsi)+\omega,\lambda-{np}/{2}\right)$.

Note that a multiple scaled matrix variate generalized hyperbolic distribution was derived by \cite{thabane04}. While the distribution they derive is sometimes referred to as a matrix variate generalized hyperbolic distribution, the model of \cite{thabane04} is in fact multiple scaled --- a fact that may be confirmed by observing that they use a matrix variate distribution for the mixing variable $\mathbf{W}$. Not only does this mean that the distribution presented by \cite{thabane04} is different to the matrix variate generalized hyperbolic distribution presented herein, but it also means that neither one of these distributions is a special case of the other. Some useful details about the multiple scaled generalized hyperbolic distribution are given by \citet[][Chp.~7]{mcnicholas16a}.

\subsection{A Matrix Variate Variance-Gamma Distribution}
We now derive the density of a matrix variate variance-gamma distribution in much the same way as the generalized hyperbolic case. However, we now take $W\sim \text{gamma}(\gamma,\gamma)$, resulting in the joint distribution 
\begin{equation*}\begin{split}
f(\vecX,w|\bvtheta)=&\frac{\gamma^{\gamma}}{(2\pi)^{\frac{np}{2}}| \matsig |^{\frac{p}{2}} |\matPsi |^{\frac{n}{2}}\Gamma(\gamma)}w^{\gamma-\frac{np}{2}-1}\\
&\times\exp\left\{-\frac{1}{2w}\tr\left(\matsig^{-1}(\vecX-\matm-w\vecA)\matPsi^{-1}(\vecX-\matm-w\vecA)'\right)-\gamma w\right\}.
\end{split}\end{equation*}
Following the same procedure as before, the density of $\fX$ is then
\begin{align*}
f_{\text{MVVG}}(\vecX|\bvtheta)=&\frac{2\gamma^{\gamma}\exp\left\{\tr(\matsig^{-1}(\vecX-\matm)\matPsi^{-1}\vecA') \right\}}{(2\pi)^{\frac{np}{2}}| \matsig |^{\frac{p}{2}} |\matPsi |^{\frac{n}{2}}\Gamma(\gamma)}  \left(\frac{\delta(\vecX;\matm,\matsig,\matPsi)}{\rho(\vecA,\matsig,\matPsi)+2\gamma}\right)^{\frac{\left(\gamma-{np}/{2}\right)}{2}} \\
&\qquad\qquad\qquad\times  K_{\left(\gamma-\frac{np}{2}\right)}\left(\sqrt{\left[\rho(\vecA,\matsig,\matPsi)+2\gamma\right]\left[\delta(\vecX;\matm,\matsig,\matPsi)\right]}\right),
\end{align*}
where $\gamma>0$. We will denote this distribution by $\text{MVVG}_{n\times p}(\matm,\vecA,\matsig,\matPsi,\gamma)$.
Note that
$W|\vecX\sim \text{GIG}\left(\rho(\vecA,\matsig,\matPsi)+2\gamma,\delta(\vecX;\matm,\matsig,\matPsi),\gamma-{np}/{2}\right)$.

\subsection{A Matrix Variate NIG Distribution}
Finally, we consider a matrix variate NIG distribution. Derived in much the same way as the previous distributions, we take $W\sim \text{IG}(1,\tgamma)$. The joint density of $\fX$ and $W$ is 
\begin{align*}
f(\vecX,w|\bvtheta)&=\frac{1}{(2\pi)^{\frac{np}{2}+1}| \matsig |^{\frac{p}{2}} |\matPsi |^{\frac{n}{2}}}w^{-\left(\frac{3+np}{2}\right)}\\&\times \exp\left\{-\frac{1}{2w}\left(\tr\left(\matsig^{-1}(\vecX-\matm-w\vecA)\matPsi^{-1}(\vecX-\matm-w\vecA)'\right)+1\right)-\frac{w\tgamma^2}{2}+\tgamma\right\},
\end{align*}
and the density of $\fX$ is then
\begin{align*}
f_{\text{MVNIG}}(\vecX|\bvtheta)&=\frac{2\exp\left\{\tr(\matsig^{-1}(\vecX-\matm)\matPsi^{-1}\vecA')+\tgamma\right\}
}{(2\pi)^{\frac{np}{2}+1}| \matsig |^{\frac{p}{2}} |\matPsi |^{\frac{n}{2}}}\left(\frac{\delta(\vecX;\matm,\matsig,\matPsi)+1}{\rho(\vecA,\matsig,\matPsi)+\tgamma^2}\right)^{-{\left(1+np\right)}/{4}}\\
&\times K_{-{(1+np)}/{2}}\left(\sqrt{\left[\rho(\vecA,\matsig,\matPsi)+\tgamma^2\right]\left[\delta(\vecX;\matm,\matsig,\matPsi)+1\right]}\right),
\end{align*}
where $\tgamma>0$. We denote this distribution by $\text{MVNIG}_{n\times p}(\matm,\vecA,\matsig,\matPsi,\tgamma)$, and note that $W|\vecX\sim \text{GIG}\left(\rho(\vecA,\matsig,\matPsi)+\tgamma^2,\delta(\vecX;\matm,\matsig,\matPsi)+1,-{(1+np)}/{2}\right)$.

\subsection{Some Properties}
One interesting element that we see for all three of these distributions is that there is a relationship between each of these matrix variate distributions and their multivariate counterparts. Specifically,
\begin{equation*}\begin{split}
\fX\sim \text{MVGH}_{n\times p}(\matm,\vecA,\matsig,\matPsi,\omega,\lambda) &\iff \vecc(\fX) \sim \text{GH}_{np}(\vecc(\matm),\vecc(\vecA),\matPsi\otimes\matsig,\omega,\lambda),\\
\fX\sim \text{MVVG}_{n\times p}(\matm,\vecA,\matsig,\matPsi,\gamma) &\iff \vecc(\fX) \sim \text{VG}_{np}(\vecc(\matm),\vecc(\vecA),\matPsi\otimes\matsig,\gamma),\\
\fX\sim \text{MVNIG}_{n\times p}(\matm,\vecA,\matsig,\matPsi,\tgamma) &\iff \vecc(\fX) \sim \text{NIG}_{np}(\vecc(\matm),\vecc(\vecA),\matPsi\otimes\matsig,\tgamma).
\end{split}\end{equation*}
These properties can be easily seen by using the representation of $\fX$ given in \eqref{eq:mmvmix} as well as the property of the matrix variate normal distribution given in \eqref{eq:normprop}. 

We can also easily derive the moment generating functions for each of these three distributions. Using the representation for a random matrix $\fX$ given in \eqref{eq:mmvmix} and the moment generating function for the matrix variate normal distribution given in \cite{Dutilleul99}, we have that the moment generating function in the general case of a matrix normal variance-mean mixture is
\begin{align*}
M_{\fX}(\vecT)&=\mathbb{E}[\exp\{\tr(\vecT'\fX)\}]
=\mathbb{E}[ \mathbb{E}[\exp\{\tr(\vecT'\fX)\}~|~W]]\\
&=\exp\{\tr(\vecT'\matm)\}\mathbb{E}[\exp\{W\tr(\vecT'\vecA+\vecT\matsig \vecT'\matPsi)\}]\\
&=\exp\{\tr(\vecT'\matm)\}M_{W}(\tr(\vecT'\vecA+\vecT\matsig \vecT'\matPsi)),
\end{align*}
where $M_{W}(\cdot)$ is the moment generating function of $W$. Therefore, in the case of the generalized inverse Gaussian distribution, we have that the moment generating function is
$$
\exp\{\tr(\vecT'\matm)\}\left[1-2\frac{\tr(\vecT'\vecA+\vecT\matsig \vecT'\matPsi)}{\omega}\right]^{-\frac{\lambda}{2}}\frac{K_{\lambda}\left(\sqrt{\omega(\omega-2\tr(\vecT'\vecA+\vecT\matsig \vecT'\matPsi))}\right)}{K_{\lambda}(\omega)}.
$$
For the variance-gamma distribution, the moment generating function is
$$
M_{\fX}^{\text{MVVG}}(\vecT)=\exp\{\tr(\vecT'\matm)\}\left(1-\frac{\tr(\vecT'\vecA+\vecT\matsig \vecT'\matPsi)}{\gamma}\right)^{-\gamma}
$$
for $\tr(\vecT'\vecA+\vecT\matsig \vecT'\matPsi)<\gamma$ and, in the case of the NIG distribution, the moment generating function is
$$
M_{\fX}^{\text{MVVG}}(\vecT)=\exp\{\tr(\vecT'\matm)\}\exp\left\{\tgamma\left(1-\sqrt{1-\frac{2\tr(\vecT'\vecA+\vecT\matsig \vecT'\matPsi)}{\tgamma^2}}\right)\right\}.
$$

Parameter estimation can be performed using expectation-conditional maximization (ECM) algorithms \citep{meng93} by treating the data as incomplete. Details are not given here but the algorithms are equivalent to one-component versions of the ECM algorithms described by \cite{gallaugher18}.

\section{Example}

We now consider a simple example for each of the three different distributions. Common elements between the distributions are as follows. We take 50 datasets each with 100 observations. For each distribution, we take
$$
\begin{array}{ll}
\matm=\left(
\begin{array}{rrrr}
 -5 & 0 & 0 & 1 \\ 
   -2 & 1 & 3 & 0 \\ 
   0 & 0 & 6 & 1 \\ 
  \end{array}
\right),&
\vecA=\left(
\begin{array}{rrrr}
1 & -1 & 0 & 1 \\ 
   0.5 & -1 & 0 & -0.5 \\ 
   0 & -1 & 0 & 0 \\ 
  \end{array}
\right).
\end{array}
$$
and the scale matrices $\matsig$ and $\matPsi$ are
$$
\begin{array}{ll}
\matsig=\left(
\begin{array}{ccc}
1 & 0.5 & 0.1 \\ 
0.5 & 1 & 0.5 \\
 0.1 & 0.5 & 1 \\
\end{array}\right),&
\matPsi=\left(
\begin{array}{cccc}
1 & 0 & 0 & 0 \\ 
0 & 1 & 0.5 & 0.5 \\
0 & 0.5 & 1 & 0.1 \\ 
0 & 0.5 & 0.1 & 1 \\
  \end{array}
\right)
\end{array}.
$$
We take the additional parameters to be $\lambda_2=-2$, $\omega=2$ for the matrix variate generalized hyperbolic, $\gamma_2=4$ for the matrix variate variance-gamma and $\tgamma_2=2$ for the matrix variate NIG distribution.
In Figure~1 (Appendix~A), we show the marginal distributions of the columns for each distribution of a typical dataset. We label the columns V1, V2, V3, and V4. The marginal location (mode) is shown by the red dashed line. 

The component-wise means and standard deviations (in brackets) of the parameter estimates are given in Table~\ref{tab:Sim2_Res}. For all three of the distributions, we get good average estimates in general. However, one unexpected outcome is the estimate for $\lambda$ for the matrix variate generalized hyperbolic distribution. The estimate is very different from the true value, and there is a very large amount of variation. We also notice a deflation, in absolute value, of the estimated entries of the skewness matrix $\vecA$ as well as a fair amount of variation. One possible explanation is that the generalized hyperbolic distribution is over-parameterized; in which case, the deflation in the estimates for the entries of $\vecA$ could be compensation for the increased value of $\lambda$.

\begin{table}[!htb]
\centering
\caption{Component-wise averages and standard deviations for the estimated parameters for each of the three distributions.}
\scalebox{0.6}{\begin{tabular}{cccccc}
\hline
\multicolumn{6}{c}{Generalized Hyperbolic}\\
\hline
$\matm$ (sd)&$\vecA$ (sd)&$\matsig$ (sd)&$\matPsi$ (sd)&$\lambda$ (sd)&$\omega$ (sd)\\
\hline
\shortstack{
$
\left[
\begin{array}{rrrr}
-4.97 & 0.05 & -0.03 & 1.02 \\ 
  -1.89 & 1.01 & 3.00 & 0.05 \\ 
  0.10 & -0.01 & 5.98 & 0.97 \\ 
\end{array}
\right]
$
\\
$
\left(\left[
\begin{array}{rrrr}
0.212 & 0.281 & 0.282 & 0.247 \\ 
  0.199 & 0.266 & 0.245 & 0.259 \\ 
  0.251 & 0.160 & 0.239 & 0.218 \\ 
\end{array}
\right]\right)
$
}
&
\shortstack{
$
\left[
\begin{array}{rrrr}
0.57 & -0.69 & 0.02 & 0.64 \\ 
  0.23 & -0.68 & -0.02 & -0.34 \\ 
  -0.02 & -0.64 & 0.04 & 0.02 \\ 
\end{array}
\right]
$
\\
$
\left(\left[
\begin{array}{rrrr}
0.526 & 0.820 & 0.272 & 0.660 \\ 
  0.276 & 0.779 & 0.255 & 0.398 \\ 
  0.338 & 0.665 & 0.173 & 0.242 \\ 
\end{array}
\right]\right)
$
}
&
\shortstack{
$
\left[
\begin{array}{rrr}
1.00 & 0.50 & 0.10 \\ 
  0.50 & 0.99 & 0.50 \\ 
  0.10 & 0.50 & 1.00 \\ 
\end{array}
\right]
$
\\
$
\left(\left[
\begin{array}{rrr}
0.000 & 0.055 & 0.061 \\ 
  0.055 & 0.117 & 0.079 \\ 
  0.061 & 0.079 & 0.112 \\ 
\end{array}
\right]\right)
$
}
&
\shortstack{
\\
\\
\\
$
\left[
\begin{array}{rrrr}
0.63 & 0.00 & 0.01 & 0.00 \\ 
  0.00 & 0.64 & 0.33 & 0.32 \\ 
  0.01 & 0.33 & 0.63 & 0.07 \\ 
  0.00 & 0.32 & 0.07 & 0.64 \\ 
  \end{array}
\right]
$
\\
$
\left(\left[
\begin{array}{rrrr}
0.606 & 0.057 & 0.068 & 0.045 \\ 
  0.057 & 0.581 & 0.299 & 0.297 \\ 
  0.068 & 0.299 & 0.596 & 0.068 \\ 
  0.045 & 0.297 & 0.068 & 0.607 \\ 
  \end{array}
\right]\right)
$
}
& 
\shortstack{
1.63
\\
(2.42)
}
& 
\shortstack{
4.08
\\
(1.33)
}
\\
\hline
\multicolumn{6}{c}{Variance-Gamma}\\
\hline
$\matm$ (sd)&$\vecA$ (sd)&$\matsig$ (sd)&$\matPsi$ (sd)&\multicolumn{2}{c}{$\gamma$ (sd)}\\
\hline
\shortstack{
$
\left[
\begin{array}{rrrr}
-4.98 & 0.01 & 0.04 & 0.96 \\ 
  -1.98 & 1.00 & 3.02 & 0.02 \\ 
  0.02 & 0.05 & 6.07 & 1.03 \\ 
\end{array}
\right]
$
\\
$
\left(\left[
\begin{array}{rrrr}
0.280 & 0.229 & 0.254 & 0.260 \\ 
  0.233 & 0.240 & 0.206 & 0.216 \\ 
  0.238 & 0.242 & 0.206 & 0.195 \\ 
\end{array}
\right]\right)
$
}
&
\shortstack{
$
\left[
\begin{array}{rrrr}
 0.98 & -0.99 & -0.00 & 1.04 \\ 
  0.49 & -0.98 & 0.01 & -0.52 \\ 
  0.00 & -1.05 & -0.06 & -0.04 \\ 
\end{array}
\right]
$
\\
$
\left(\left[
\begin{array}{rrrr}
0.307 & 0.269 & 0.256 & 0.282 \\ 
  0.248 & 0.256 & 0.222 & 0.247 \\ 
  0.260 & 0.245 & 0.232 & 0.225 \\ 
\end{array}
\right]\right)
$
}
&
\shortstack{
$
\left[
\begin{array}{rrr}
1.00 & 0.51 & 0.10 \\ 
  0.51 & 1.01 & 0.51 \\ 
  0.10 & 0.51 & 1.02 \\ 
\end{array}
\right]
$
\\
$
\left(\left[
\begin{array}{rrr}
0.000 & 0.048 & 0.063 \\ 
  0.048 & 0.095 & 0.081 \\ 
  0.063 & 0.081 & 0.129 \\ 
\end{array}
\right]\right)
$
}
&
\shortstack{
\\
\\
\\
$
\left[
\begin{array}{rrrr}
0.99 & -0.01 & -0.01 & 0.00 \\ 
  -0.01 & 0.98 & 0.47 & 0.51 \\ 
  -0.01 & 0.47 & 0.98 & 0.09 \\ 
  0.00 & 0.51 & 0.09 & 1.00 \\ 
  \end{array}
\right]
$
\\
$
\left(\left[
\begin{array}{rrrr}
0.121 & 0.064 & 0.053 & 0.060 \\ 
  0.064 & 0.103 & 0.074 & 0.072 \\ 
  0.053 & 0.074 & 0.121 & 0.059 \\ 
  0.060 & 0.072 & 0.059 & 0.126 \\ 
  \end{array}
\right]\right)
$
}
&
\multicolumn{2}{c}{ 
\shortstack{
4.20
\\
(1.04)
}}\\
\hline
\multicolumn{6}{c}{Normal Inverse Gaussian}\\
\hline
$\matm$ (sd)&$\vecA$ (sd)&$\matsig$ (sd)&$\matPsi$ (sd)&\multicolumn{2}{c}{$\tgamma$ (sd)}\\
\hline
\shortstack{
$
\left[
\begin{array}{rrrr}
-5.02 & 0.04 & 0.01 & 1.03 \\ 
  -1.99 & 1.04 & 2.99 & 0.05 \\ 
  0.02 & 0.01 & 5.98 & 1.01 \\ 
\end{array}
\right]
$
\\
$
\left(\left[
\begin{array}{rrrr}
0.143 & 0.134 & 0.133 & 0.137 \\ 
  0.137 & 0.123 & 0.140 & 0.117 \\ 
  0.148 & 0.120 & 0.128 & 0.114 \\ 
\end{array}
\right]\right)
$
}
&
\shortstack{
$
\left[
\begin{array}{rrrr}
1.16 & -1.18 & 0.01 & 1.02 \\ 
  0.55 & -1.19 & 0.04 & -0.64 \\ 
  0.01 & -1.11 & 0.04 & 0.02 \\ 
\end{array}
\right]
$
\\
$
\left(\left[
\begin{array}{rrrr}
0.506 & 0.446 & 0.306 & 0.418 \\ 
  0.390 & 0.462 & 0.323 & 0.357 \\ 
  0.298 & 0.433 & 0.271 & 0.249 \\ 
\end{array}
\right]\right)
$
}
&
\shortstack{
$
\left[
\begin{array}{rrr}
1.00 & 0.49 & 0.11 \\ 
  0.49 & 1.01 & 0.51 \\ 
  0.11 & 0.51 & 1.00 \\ 
\end{array}
\right]
$
\\
$
\left(\left[
\begin{array}{rrr}
0.000 & 0.045 & 0.053 \\ 
  0.045 & 0.107 & 0.077 \\ 
  0.053 & 0.077 & 0.119 \\ 
\end{array}
\right]\right)
$
}
&
\shortstack{
\\
\\
\\
$
\left[
\begin{array}{rrrr}
1.02 & 0.01 & 0.01 & 0.02 \\ 
  0.01 & 1.06 & 0.54 & 0.53 \\ 
  0.01 & 0.54 & 1.06 & 0.11 \\ 
  0.02 & 0.53 & 0.11 & 1.07 \\ 
  \end{array}
\right]
$
\\
$
\left(\left[
\begin{array}{rrrr}
0.250 & 0.065 & 0.064 & 0.072 \\ 
  0.065 & 0.285 & 0.175 & 0.139 \\ 
  0.064 & 0.175 & 0.281 & 0.072 \\ 
  0.072 & 0.139 & 0.072 & 0.245 \\ 
  \end{array}
\right]\right)
$
}
&
\multicolumn{2}{c}{ 
\shortstack{
2.12
\\
(0.50)
}}\\
\hline
\end{tabular}}
\label{tab:Sim2_Res}
\end{table}

\section{Discussion}
We derived the densities and described parameter estimation for three matrix variate skew distributions using a matrix normal variance-mean mixture. The three distributions were the matrix variate generalized hyperbolic, variance-gamma and NIG distributions, respectively. When looking at the estimates in the simulations, we obtained fairly good results. One exception was the average estimates of $\lambda$ and the skewness matrix $\vecA$ for the matrix variate generalized hyperbolic distribution. However, this could be due to over-parameterization. 

One possible extension of the work herein is to consider multiple-scaled analogues of the matrix variate variance-gamma, NIG and skew-$t$ distributions. The resulting multiple scaled distributions would be arrived at in an analogous fashion to the multiple-scaled matrix variate generalized hyperbolic distribution of \cite{thabane04}. Finally, it would be interesting to consider placing a constrained covariance structure on $\matsig$ for possible use with multivariate longitudinal data, i.e., data where multiple quantities are measured over time.

{\small\section*{Acknowledgements}
The authors are grateful for the support of the Vanier Canada Graduate Scholarships (Gallaugher) and the Canada Research Chairs (McNicholas) programs. 

 \newpage
 \appendix
 \section{Figure}\label{sec:app}
\begin{figure}[!htb]
\centering
\includegraphics[width=0.625\textwidth]{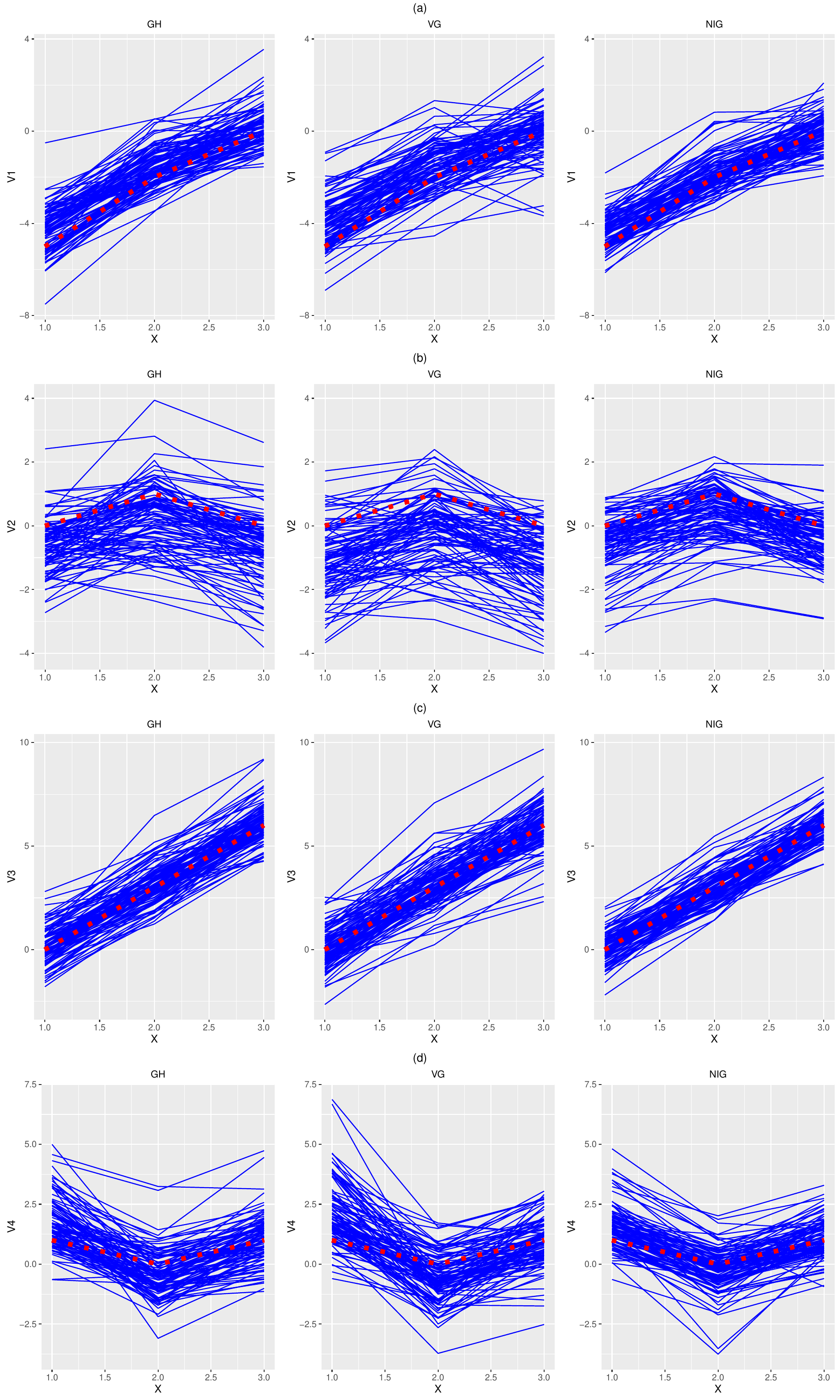}
\caption{Marginal distributions for the matrix variate GH, VG and NIG distributions for (a) V1, (b) V2, (c) V3 and (d) V4. The marginal location is by a red dashed line.}
\label{fig:Sim2}
\end{figure}}

\end{document}